\documentclass[twocolumn,aps,pra,showpacs]{revtex4}
\usepackage{epsfig}
\begin{document}
\title{Noise excess free record/upload of non-classical states to continuous-variable quantum memory}
\author{Radim Filip}
\affiliation{Department of Optics, Palack\' y University,\\
17. listopadu 50,  772~07 Olomouc, \\ Czech Republic}
\date{\today}
\begin{abstract}
In recent continuous-variable quantum memory experiment, one quadrature of light pulse is directly uploaded by the light-atom coupling whereas complementary quadrature is obtained by homodyne measurement of the out-coupled light. Subsequently, information from homodyne measurement is written into the memory by feed-back electro-optical control of the atomic state. Using the same experimental set-up, a deterministic noise excess free record of unknown quantum states to continuous-variable quantum atomic memory is proposed. Further, the memory experiment is extended by the pre-squeezing of the recorded state of light and squeezing post-correction of the recorded state. To upload a resource to the memory, as single-photon state or superposition of coherent states, the post-selection of the measurement results from homodyne detection is suggested. Then such the probabilistic upload approaches even a lossless transfer of these highly nonclassical states into the quantum memory.
\end{abstract}

\maketitle

A non-classicality of quantum states is extremely fragile to disturbing effects of a noise coupled to the system. This is a real problem for the transmission and operations with the non-classical states. Recently, the same problem arises also in a construction of quantum memory. The quantum memories have been recently demonstrated using electromagnetically-induced transparency \cite{QM1} and off-resonant Faraday effect \cite{memory}. It is useful to distinguish three basic steps: initial record (or upload)
of the quantum state, its storage for a longer time inside the memory and final read-out from the memory. In addition, quantum state written to the memory can either carry encoded information or it can be a resource for other quantum operations. To distinguish these two cases, the record of quantum states is used for the states carrying information and the upload of quantum state for the resources. In the former case, encoded information makes the state to be at least partially unknown, whereas in the latter case, the uploaded state is known completely and the upload can be optimized with a respect to that specific state. In this case, a relatively fast probabilistic upload is sufficient to transfer the resource state into the memory.

The quantum memory should definitely preserve the non-classical properties of quantum states. It is a necessary for any application of quantum memory, for example, for the quantum repeaters for the long-distance quantum key distribution. Focusing on the record/upload process, the noise during this process has to be investigated. There is always vacuum noise which accompanies any loss in the record/upload, but even more destroying is an impact of a noise excess above this vacuum noise penalty. Therefore, an noise excess free record/upload to quantum memory should be find. In the other words, the noise excess free means that the record/upload is only pure lossy process. Such the record/upload will be not entanglement breaking channel \cite{Lee}, will not completely vanish sub-Poissonian statistics of single photon state \cite{Mandel} and also will not smear out interference effects in the superposition of coherent states \cite{Walls}. Also security of the continuous-variable quantum key distribution is preserved for any loss in the transfer of quantum state if there is no noise excess \cite{GG}. Second, more complicated open problem is how to construct the lossless upload of the nonclassical state. Such the upload allows to prepare highly non-classical inside the memory.

The quantum memory for coherent states of light has been recently demonstrated using a pair of glass cells filled by Cs atoms (at room temperature) \cite{memory}. In this experiment, the coherent state is generated by the amplitude and phase modulations of the strong continuous-wave (cw) millisecond laser pulses. The off-resonant interaction of cw light pulses with the collective spin atoms inside the cell (Faraday effect) can be describe as quantum non-demolition coupling between effective quadratures of the light and atomic modes. The coupling automatically records a single quadrature of light to the effective quadrature of collective spin of the atoms. The complementary quadrature of light pulse passing through the memory is then measured by homodyne detection and feed-forward by the electro-optical control into the atomic memory. In the experiment \cite{memory}, an average fidelity of recorded coherent states has been shown to be higher than a threshold corresponding to measure-prepare strategy. To approach unit fidelity in the same set-up, a difficult high squeezing of the collective spin state of the atoms before the uploading procedure is required. Since the average fidelity has been considered as a figure of merit, the demonstrated record of the coherent state is not the excess noise free and the recorded state has been far a way from a pure state.

There were few attempts to achieve noise excess free memory, involving several coherences of each atom in the cell \cite{Opatrny05} or multi-pass version of the deterministic memory \cite{Sherson06}. Further, the scheme with two passes has been theoretically examined with an additional magnetic field to achieve the deterministic and noise excess free memory with a loss exponentially decreasing as the coupling constant is growing \cite{Muschik06}. These experimentally challenging extensions are mainly orientated to approach the deterministic record/read-out of unknown quantum state carrying information into/from memory. On the other hand, to upload the known highly non-classical resource state, such as the single photon state or the superposition of coherent states into the memory, the probabilistic approach could be more simple and effective.

In this paper, the noise excess free version of deterministic record of unknown quantum state into the quantum memory is described and the probabilistic upload of highly non-classical states (single-photon state, superposition of coherent states) approaching lossless light-atomic transfer is proposed.

In the Section II., is shown that the original set-up from Ref.~\cite{memory} can be directly used to obtain noise excess free deterministic record of any quantum state of light for an arbitrary weak non-demolition coupling and without any noise pre-squeezing of the atomic memory. The recorded state is noise excess free but only up to a squeezing of the collective atomic variables. Therefore, a squeezing post-correction is suggested to approach phase-insensitive record symmetrical in both the quadratures. Further, it is demonstrated that a pre-squeezing of the state of light before entering into the memory remarkably enhances the coupling strength and subsequently, it reduces the attenuation of the uploaded state inside the memory. Unfortunately, the application of the squeezing post-correction actively is always at a cost of additional loss and for a higher squeezing of the recorded state in the memory, the additional attenuation is larger. In the result, the purely lossy phase-insensitive record with maximum $1/4$ of transmission can be achieved without any noise excess. It can be restrictive for the upload of the non-classical resources, such as the single-photon state or the superposition of the coherent states. If the upload is lossy, such the single photon is uncontrollably transferred to the memory, without any trigger saying whether the photon is really uploaded.

Therefore, in the Section III., a probabilistic upload is proposed based on a post-selection of a result from the homodyne detector in the same memory set-up. The single photon can be probabilistically uploaded into the memory without loss and with an arbitrary precision, although the coupling between the light and atoms in the memory can be weak. The superposition of coherent states can be probabilistically uploaded with a reduced amplitude, but with the purity approaching unity. Remarkably, for the probabilistic upload the squeezing post-correction can be omitted if the light-atom coupling is sufficiently weak. To increase the success rate of the single-photon upload, the pre-squeezing of the state of light before the upload can be used. The superposition of coherent states can be uploaded in the same way, but the pre-squeezing plays here more important role, it effectively enhances the amplitude of the coherent states in the uploaded superposition, resulting in more distinguishable interference of the coherent states. Because all these proposals are within the existing experiment set-up, they can be directly implemented if the suitable non-classical sources are available.

\section{Noise excess free deterministic record into quantum memory}

The noise excess free (phase-insensitive) quantum record/upload of single-mode light into the atomic memory can be defined by the transformation
\begin{equation}\label{noiseless}
X'_A=\sqrt{T}(X_L+X_N),\nonumber\\
P'_A=\sqrt{T}(P_L+P_N)
\end{equation}
of the light quadratures $X_L$ and $P_L$ to the atomic quadratures $X_A$ and $P_A$ in the Heisenberg picture. Here $T$ stands for the transmission coefficient of the record/upload. The quadrature noisy operators $X_N$ and $P_N$ describe the added noises in the state transfer. The variance of vacuum noise is considered to be unity. From the Heisenberg uncertainty principle, the minimal variance of both $X_N$ and $P_N$ is $V_N=1-T$  ($T\leq 1$). In this case, there is no noise excess in the record/upload process. It corresponds to transmission exhibiting only a pure loss, which can be modeled by a virtual beam splitter between light and atoms with vacuum in the free port. Such the noise excess free record/upload is advantageous because it can preserve some important non-classical properties of quantum states. It will never completely break entanglement of Gaussian state, vanish sub-Poisson behavior of single-photon state \cite{Mandel} or quantum superposition of coherent states \cite{Walls}. Also only loss in the record/upload will not break security of the continuous-variable key distribution protocol with coherent states if the reverse reconciliation is used \cite{GG}.

Basically, information encoded into the quadratures and Gaussian entanglement or security of key distribution are not changed if the recorded/upoload state is only unitarily transformed by a known Gaussian operation like the phase shift, displacement and squeezing. In this case, the record/upload is noise excess free up to that unitary operation. Particularly, we focus on the noise excess free record/upload up to the squeezing operation: $X'_A\rightarrow aX'_A$, $P'_A\rightarrow P'_A/a$. For an application, either the squeezing could be actively post-corrected in the memory, as will be proposed below, or if it is not necessary, it can be finally simply corrected on the measured data. To increase a quality of the upload, also a pre-squeezing operation $X_L\rightarrow X_L/c$, $P_L\rightarrow cP_L$ on the light mode before record/upload can be considered. It transforms the input state to its squeezed version before the record/upload into the memory. It will be shown, both the pre-squeezing of input state and the post-squeezing correction can remarkably help to reach the noise excess free record or even the lossless upload of quantum state into the memory.

Consider now the quantum memory experiment in Ref.~\cite{memory}, see also Ref.~\cite{Bruss} for more details. In that experimental setup, there are two simultaneously available quantum non-demolition (QND) interactions between the light and atoms inside the cells. Basically, they couple together the mode of light described by two complementary quadratures $X_L$ and $P_L$ and the effective collective atomic mode having complementary quadratures $X_A$ and $P_A$. Both the QND transformations can be simply described in Heisenberg picture:
\begin{eqnarray}\label{coupl1}
X'_L=X_L+\kappa P_A,\,\,\, P'_L=P_L,\nonumber\\
X'_A=X_A+\kappa P_L,\,\,\, P'_A=P_A,
\end{eqnarray}
and
\begin{eqnarray}\label{coupl2}
X'_L=X_L+\kappa X_A,\,\,\, P'_L=P_L,\nonumber\\
X'_A=X_A,\,\,\, P'_A=P_A-\kappa P_L,
\end{eqnarray}
where $\kappa$ is an effective coupling constant \cite{memory}. Either coupling (\ref{coupl1}) or (\ref{coupl2}) can be separately activated in the same set-up \cite{Bruss}. Both are the particular QND interactions, but the following analysis is generally valid for any kind of the QND transformation between the quadratures of light and atomic memory.

If the coupling (\ref{coupl1}) is considered, the quadrature $P_L$ can be directly written into the memory by the light-atom interaction, up to the added noise from the atomic quadrature $X_A$. To write the complementary quadrature $X_L$, the light pulse passing through the memory is measured by homodyne detection and the photocurrent controls the magnetic field applied with an adjustable gain to the atomic cells. By this feed-forward technique, the atomic quadrature $P_A$ can be displaced whereas the quadrature $X_A$ is not disturbed. This is standard record mechanism used in the Ref.~(\cite{memory}). Now we assume the record of unknown quantum state up to the squeezing of the recorded state. After such the procedure, the transformation of the atomic quadratures takes the following form
\begin{eqnarray}\label{record}
X_A^{'out} &=& a(X_A+\kappa P_L^{in}),\nonumber\\
P_A^{'out} &=& \frac{1}{a}\left(P_A+g(X_L^{in}+\kappa
P_A)\right)=\nonumber\\
& &\frac{1}{a}\left((1+g\kappa)P_A+gX_L^{in}\right),
\end{eqnarray}
where $g$ is an overall gain of the feed-forward correction and $a$ is a scaling factor representing the squeezing of recorded state. The effective atomic mode is considered initially in vacuum state with the unit variance. To obtain the noise excess free record up to the squeezing, the parameters have to satisfy the following equations
\begin{eqnarray}\label{eq1}
a^2\kappa^2=1-a^2,\nonumber\\
\frac{g^2}{a^2}=1-\frac{(1+g\kappa)^2}{a^2}.
\end{eqnarray}
The solution of (\ref{eq1}): $g=-\kappa/(1+\kappa^2)$ and $a=1/\sqrt{1+\kappa^2}$,
gives a possibility to achieve the noise excess free record described by Eqs.~(\ref{noiseless}) up to the
squeezing. The record transmission coefficient
\begin{equation}
T_R=\frac{\kappa^2}{1+\kappa^2}
\end{equation}
of the transfer from the light mode to atomic mode shows that an unknown state written inside the memory is always squeezed by
the factor $a=1/\sqrt{1+\kappa^2}$. But even for very small $\kappa>0$, any state can be (up to the squeezing) written into the memory with no noise excess.
For a feasible gain around $\kappa=1$, the upload can be up to the noise excess free attenuation $T_R=0.5$ up to the squeezing. 

Such the noise excess free record even without the squeezing correction can be useful, for example, for the manipulations with Gaussian entanglement. Using computable measure of entanglement for Gaussian state \cite{logneg}, it is straightforward to prove the following comparison. Let us compare the cases after the record, without squeezing correction and with perfect squeezing correction. Consider single mode from two-mode Gaussian entangled state which is recorded into the memory. For any single-mode noisy Gaussian operation after the noise excess free record (up to the squeezing), even with an arbitrary small $\kappa>0$, no matter that the uncorrected squeezing reduces entanglement it will never help to subsequent Gaussian channel to completely break entanglement. It means that for both the cases, the thresholds to pass entanglement through the record procedure plus consecutive Gaussian channel are identical. It illustrates a practical advantage of the noise excess free record up to the squeezing. Such the result is generally impossible to obtain if there is a noise excess in the record.

To actively compensate the squeezing of the recorded state inside the memory, the squeezing post-correction is necessary. The squeezing post-correction should make transformation $X''_A=X'_A/b$ and $P''_A=bP'_A$ to reduce effect of the squeezing in the atomic memory. Since the total record process could exhibit pure loss, the squeezing post-correction can be implemented at a cost of an additional loss in the record.
To build post-correction operation, the second kind of the coupling (\ref{coupl2}) with the coupling constant $\kappa'$ is now assumed in the same experimental set-up. Now, the light mode described by $X_L^0$ and $P_L^0$ is considered in vacuum state. Thus only bright local oscillator is injected into the cells. The homodyne measurement of $X'_L$ quadrature of the pulse outgoing from the cells is followed by the feed-forward displacement of the atomic quadrature $X_A$ by the magnetic field. To achieve a desirable squeezing post-correction $X''_A=X'_A/b$ and $P''_A=bP'_A$ ($b<1$ is considered to compensate the squeezing in Eqs.~(\ref{record})) up to a pure loss (but without any noise excess), the transformations
\begin{eqnarray}
P_A^{''out} &=& bP_A^{'in}-\kappa' P_L^0,\nonumber\\
X_A^{''out} &=&\frac{1}{b}(1+g\kappa')X^{'in}_A+gX_L^0
\end{eqnarray}
have to satisfy the following conditions
\begin{eqnarray}
b^2=1-\kappa'^2,\nonumber\\
\frac{(1+g\kappa')^2}{b^2}=1-g^2.
\end{eqnarray}
It can be achieved if $g=-\kappa'$ and then
$b^2=1-\kappa'^2$ is given for any $\kappa'<1$. The additional losses introduced by the
post-correction are characterized by the transmission $T_C=b^2=1-\kappa'^2<1$. Since $b<1$, the squeezing post-correction can be performed only in the $P_A$ quadrature. To precisely compensate the squeezing introduced in the
record, it is necessary to adjust $a^2=b^2$. From this follows that
$\kappa'^2=\kappa^2/(1+\kappa^2)$ and total transmission coefficient of the record plus squeezing correction is given by
\begin{equation}
T=T_CT_L=\kappa^2\frac{1-\kappa'^2}{1+\kappa^2}=\frac{\kappa^2}{(1+\kappa^2)^2}.
\end{equation}
The maximal transmission $T_{max}=1/4$ is achieved for $\kappa=1$ ($\kappa'=1/\sqrt{2}$) which is feasible in the experiment in Ref.~\cite{memory}. As the result, any quantum state can be in principle deterministically recorded without noise excess as is described by Eq.~(\ref{noiseless}) up to the transmission $T=1/4$.

Another open possibility how to compensate the squeezing in the recorded state is a pre-squeezing $P_L\rightarrow c P_L$ and $X_L\rightarrow X_L/c$ of the state which should be recorded. From a simple calculation follows that this configuration does not lead in principle to the noise excess free record. But if both the pre-squeezing and squeezing post-correction are considered together, the local pre-squeezing helps to increase the coupling constant $\kappa$. Taking both the pre-squeezing and squeezing post-correction into the consideration, the record mechanism is described by the transformations:
\begin{eqnarray}
X_A^{'out} &=& a(X_A+\kappa ),\nonumber\\
P_A^{'out} &=& \frac{1}{a}\left(P_A+g(\frac{1}{c} X_L^{in}+\kappa
P_A)\right)=\nonumber\\
& &\frac{1}{a}\left((1+g\kappa)P_A+\frac{g}{c}X_L^{in})\right),
\end{eqnarray}
where $c$ stands for the scaling factor corresponding to the pre-squeezing of the light mode. To approach the noise excess free record, it is necessary
to adjust electronic gain $g=-c^2\kappa/(1+c^2\kappa^2)$ and then the squeezing
correction after the record is determined by $a^2=1/(1+c^2\kappa^2)$.
Thus for a given $\kappa$, the squeezing post-correction has to be larger ($a$ should be smaller) as $c$ is bigger. Interestingly, the transmission coefficient
\begin{equation}
T'_{R}=\frac{1}{1+\frac{1}{\kappa^2c^2}}
\end{equation}
of the record up to the squeezing is effectively increased as the squeezing factor $c$ is larger. Remarkably, by only local pre-squeezing of the state, which should be recorded into memory, the coupling constant $\kappa$ is increased. Since $T'_R$ approaches unity for sufficiently large $c$, a lossless record up to the squeezing can be approached. It is even possible for an arbitrary small $\kappa>0$ if the sufficient pre-squeezing is available. On the other hand, if the uploaded state inside the memory should be not squeezed, the squeezing correction has to be applied. Then, to reach the limit $T_{max}=1/4$, $c\kappa=1$ has to be satisfied and it can be approached for any $\kappa$ as the pre-squeezing is increasing.

This deterministic record can be used to upload not only coherent states and Gaussian entangled states, but also highly nonclassical single-photon state as an useful quantum resource.
The single-photon state can be prepared by the single-photon subtraction from the squeezed light \cite{photon}. Actually, the state prepared by such the procedure approaches the squeezed single-photon state \cite{Fiurasek}. This squeezing would be advantageous for the upload procedure as has been discussed. But, to have exactly single-photon state inside memory, it is necessary to actively apply the squeezing post-correction. This operation can be naturally included into the uploading scheme
proposed above. In a result, to get single-photon state inside the memory, the single-photon subtraction can be
simply applied on the squeezed state followed by the uploading procedure proposed above.
But still the upload has some uncertainty, we do not know if the
single photon state is actually inside the memory. Therefore, the uploaded state is a mixture
\begin{equation}\label{mix}
\rho=T_{tot}|1\rangle_A\langle 1|+(1-T_{tot})|0\rangle_A\langle 0|,
\end{equation}
where
\begin{equation}
T=T_CT'_R=\frac{c^2\kappa^2}{(1+c^2\kappa^2)^2},
\end{equation}
which optimally approaches $\rho=1/4|1\rangle_A\langle 1|+3/4|0\rangle_A\langle 0|$ as $c\kappa=1$. The Wigner function of that state is never negative since $T<1/2$ \cite{Ou}, but still any mixture (\ref{mix}) with $T>0$ is a non-classical state \cite{Lvovsky}. For example, the statistics of (\ref{mix}) is always sub-Poissonian for arbitrary $T>0$. In experiments with the single photons, this method can be used only if the post-selection on the fixed number of photons is finally applied to have complete control how many photons participated in the experiment. Note, there is a similar requirement for the linear optical experiments with the single photons. They work correctly only if all the photons are finally detected.

\section{Lossless probabilistic upload}

\subsection{Single-photon state}

Fortunately, there is a better way how to upload the single-photon state. Instead of the deterministic electro-optical feed-forward correction, a proper post-selection of measured data from the homodyne detector is applied. It makes the upload only probabilistic, but since the uploaded state is resource, it is acceptable if the rate is not too small. For the upload of the Gaussian states, there is no gain from this probabilistic method over the previous deterministic feed-forward correction. But for the non-classical non-Gaussian states, for example, for the squeezed single-photon state produced by the single-photon subtraction \cite{Fiurasek} the post-selection of the measured quadrature $X'_L$ in a tiny interval $\langle -B,B\rangle$ around value $X'_L=0$ will remarkably help.

To simply understand why it will help, the unitary coupling (\ref{coupl2}) between light and atoms can expand for very small $\kappa \tau\ll 1$, where $\tau$ is an interaction time, as $U_{QND}\approx 1+\kappa/4 \tau(a_A+a^{\dagger}_A)(a^{\dagger}_L-a_L)$, where $a_L=X_L+iP_L$ and $a_A=X_A+iP_A$ are annihilation operators of the light and atomic modes. In this convention, the vacuum noise has variance $1/4$. The ideal projection on the eigenstate $|x=0\rangle$ of $X'_L$ corresponding to eigenvalue $X'_L=0$ can be rewritten in a form of $|x=0\rangle\propto\sum_n|2n\rangle$. Initially, light is in the Fock state $|1\rangle_L$ and atomic mode is in the vacuum state $|0\rangle_A$. Straightforwardly, we can prove that
\begin{eqnarray}
\langle x=0|_L\left(1+\frac{\kappa}{4} \tau(a_A+a^{\dagger}_A)(a^{\dagger}_L-a_L)\right)|1\rangle_L|0\rangle_A=\nonumber\\
-\frac{\kappa}{4}|1\rangle_A
\end{eqnarray}
and from it follows that in the limit of weak coupling and very tight post-selection $(B\rightarrow 0)$, it is possible to upload single-photon state to the memory without any loss and noise.

To calculate the upload for a finite $B$, we use the formalism of Wigner functions.
The squeezed single-photon state is represented by the Wigner function
\begin{eqnarray}\label{1photon}
W_L(x,p)=\frac{2}{\pi}\exp(-\frac{2}{a^2}x^2_L-2a^2p^2_L)\times\nonumber\\
\left(4(\frac{1}{a^2}x^2_L+a^2p^2_L)-1\right),
\end{eqnarray}
where $a$ is squeezing parameter. The atomic memory is considered initially in vacuum state described by the Wigner function $W_A(x_A,p_A)=2/\pi\exp(-2x^2_A-2p^2_A)$. The application of the coupling (\ref{coupl2}) on the product $W_L(x_L,p_L)W_A(x_A,p_A)$ followed by the homodyne detection and above mentioned post-selection will generate the Wigner function
\begin{eqnarray}
W'_A(x,p)=\frac{a^3(d^2+4\kappa^2p^2)}{\pi S d^5}
\exp\left(-2\frac{a^2p^2}{d^2}-2x^2\right)\times\nonumber\\
\left(\mbox{Erf}\left[\frac{\sqrt{2}}{a}\left(B-\kappa x\right)\right]+\mbox{Erf}\left[\frac{\sqrt{2}}{a}
\left(B+\kappa x\right)\right] \right)-\nonumber\\
\frac{4\sqrt{2}}{\sqrt{\pi^3} S d}
\exp\left(-2\frac{B^2}{a^2}-2x^2\frac{d^2}{a^2}-2p^2\frac{a^2}{d^2} \right)\times\nonumber\\
\left(B\mbox{Cosh}\left[\frac{4B\kappa x}{a^2}\right]+\kappa x \mbox{Sinh}\left[\frac{4B\kappa x}{a^2}\right]\right)\nonumber\\
\end{eqnarray}
of the state uploaded into the memory,
where $d=\sqrt{a^2+\kappa^2}$ and $\mbox{Erf}[x]$ is error function. The success rate of the upload is
\begin{equation}
S=\mbox{Erf}\left[\sqrt{\frac{2}{d^2}}B\right]-\sqrt{\frac{2}{\pi}}
\exp\left(-\frac{2B^2}{d^2}\right)\frac{2a^2B}{d^3}.
\end{equation}
If the squeezing post-correction $x\rightarrow \frac{a}{d}x$ and $p\rightarrow \frac{d}{a}p$ is applied, then as $B$ goes to zero, the Wigner function of atomic state approaches
\begin{eqnarray}
\lim_{B\rightarrow 0} W'_A(x,p)&=&W_1(x,p),\nonumber\\
W_1(x,p)&=&\frac{2}{\pi}\exp(-2x^2-2p^2)\left(4(x^2+p^2)-1\right),\nonumber\\
\end{eqnarray}
where $W_1(x,p)$ is exactly the Wigner function of the single photon.
The squeezing post-correction does not depend on the post-selection threshold $B$. To approach controllable upload, the squeezing post-correction proposed in the previous section cannot be used, since it can work only up to additional losses which destroys the desired control of upload. But, fortunately, if both $B,\kappa$ are small then the squeezing post-correction can be approximately omitted and the uploaded state is
\begin{equation}
W'_A(x,p)\approx \left(W_1(x,p)+\mbox{O}^2(\kappa)\right)+\mbox{O}^2(B).
\end{equation}
As the result, the single-photon state is probabilistically uploaded into the memory with full control and remarkably, is not attenuated comparing to the previous deterministic case. Thus the post-selection compensates the loss which occurs in the deterministic upload. Since the probabilistic upload is obtained for a small $\kappa$, the post-selection also compensates strength of the coupling between light and atoms. But as $\kappa$ decreases, the post-selection threshold $B$ has to decrease and it reduces the success rate of the upload.

To simply measure quality of the single-photon state uploaded into the memory versus the success rate of the upload, the fidelity and negativity of Wigner function can be used. The fidelity between the uploaded state described by the Wigner function $W'_A(x,p)$ and target single-photon state with the corresponding Wigner function $W_1(x,p)$ is defined as
\begin{equation}\label{fi}
F=\pi\int\!\!\!\int_{-\infty}^{\infty}W'_A(x,p)W_1(x,p)dxdp,
\end{equation}
whereas the negativity in the origin of the phase space is given simply by $N=W'_A(0,0)$. The expression for the fidelity is rather complex but for both the $\kappa$ and $B$ small, the expansion
\begin{equation}\label{fid}
F\approx 1+\mbox{O}^3(\kappa)-\left(\frac{4}{3\kappa^2}+\frac{\kappa^2}{2a^4}+\mbox{O}^3(\kappa)\right)B^2+\mbox{O}^3(B)
\end{equation}
shows how the result of a sufficiently tight post-selection practically does not depend on a small $\kappa$. Without the squeezing post-correction, the maximal fidelity is exactly
\begin{equation}
F_{max}=\frac{8a^3\sqrt{(a^2+\kappa^2)^3}}{(2a^2+\kappa^2)^3},
\end{equation}
in the limit $B\rightarrow 0$.
For a small $B$, the negativity of Wigner function is
\begin{equation}
N\approx -\frac{2}{\pi}+\mbox{O}^3(\kappa)+\left(\frac{16}{3\pi\kappa^2}+\frac{4\kappa^2}{\pi a^4}+\mbox{O}^3(\kappa)\right)B^2+\mbox{O}^3(B).
\end{equation}
The maximum of negativity $N_{max}=-\frac{2}{\pi}$
can be always approached if $B$ is sufficiently small. The $N_{max}$ is independent on $a$ and $\kappa$ because the negativity of Wigner function is not changed by the pre-squeezing. Both the fidelity and negativity can be achieved with the success rate approaching
\begin{equation}
P_S\approx \left(\frac{2\sqrt{2}\kappa^2}{\sqrt{\pi}a^3}+\mbox{O}^3(\kappa)\right)B+\mbox{O}^{3}(B),
\end{equation}
which grows as pre-squeezing parameter $a$ is smaller. Here the pre-squeezing plays also positive role, it significantly enhances the probability of success. The quantitative results of the single photon upload are depicted in Fig.1. The pre-squeezing $a<1$ can evidently help to increase the success rate for both the fidelity as well as for the negativity, three curves are depicted for $a=1,0.5,0.25$. Thus squeezing of the single-photon state from the squeezed light by the single photon subtraction is actually an advantage for the upload into the memory. Contrary to the deterministic case, the post-selection can effectively increase the coupling between light and atoms.

To approach such the lossless upload of single photon state, the in-coupling and out-coupling losses have to be minimized. The in-coupling losses will mix the single-photon state with vacuum state and in a result, a less sub-Poissonian state is actually uploaded. The efficiency of homodyne detector can be can be very close to unity and also electronic noise of homodyne detection can be negligible, if the power of local oscillator is high. The out-coupling losses can be joined with the efficiency of the homodyne detection to an effective total out-coupling efficiency $\eta$. The fidelity (\ref{fi}) is a decreasing function of $\eta$, as can be seen from Fig.~2. To approach the losses upload, the coupling constant $\kappa$ has to be reasonably high and then, the pre-squeezing in the opposite direction ($a>1$) can help to reach unit fidelity of the single-photon state at a cost of the success rate.

\begin{figure}
\centerline{\psfig{width=6cm,angle=0,file=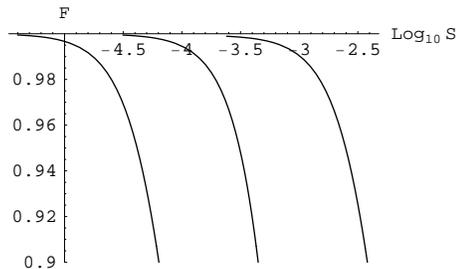}}
\caption{Fidelity $F$ of uploaded squeezed single photon state (without squeezing post-correction) with respect to ideal single-photon state in dependency on the success rate $\log_{10}S$: $\kappa=0.05$ and $a=1,0.5,0.25$ (from left to right).}
\end{figure}
\begin{figure}
\centerline{\psfig{width=5cm,angle=0,file=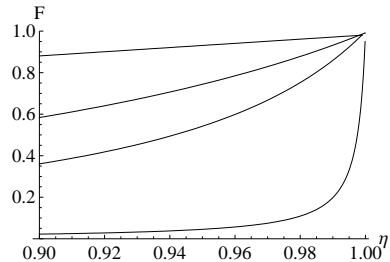}}
\caption{Fidelity $F$ of uploaded squeezed single photon state (without squeezing post-correction) with respect to ideal single-photon state in dependency on the total efficiency of homodyne detection $\eta$: $B=0.01$, $a=2,2,2,4$ and $\kappa=0.1,0.3,0.8,2$ (from bottom to top).}
\end{figure}

\subsection{Superposition of coherent states}

A superposition of the coherent states is considered as an another resource for the highly non-linear quantum operations and quantum computing \cite{Ralph}. It is rather a superposition of many Fock states, than just single Fock state, therefore it can be interesting to investigate whether the upload can transfer such the state of light into the memory. Let us consider the following superposition
\begin{equation}\label{cat}
|\Psi\rangle=\frac{1}{\sqrt{\pi(1+\exp(-2|\alpha|^2))}}
\left(|\alpha\rangle+|-\alpha\rangle\right)
\end{equation}
of coherent states which is pre-squeezed before the upload, where $\alpha\in \mbox{R}$. Then the state can be described by the Wigner function
\begin{eqnarray}
W_{L}(x,p)&=&\frac{\exp(-\frac{p^2}{a^2})}{\pi(1+\exp(-2|\alpha|^2))}\nonumber\\
& &\left[\exp\left(-2(x a+x_0)^2\right)+\exp\left(-2(x a-x_0)^2\right)+\right.\nonumber\\
& &\left. 2\exp(-2a^2x^2)\mbox{Cos}\left(\frac{4x_0p}{a}\right)\right],
\end{eqnarray}
where $x_0=\mbox{Re}(\alpha)$. At the beginning, the vacuum state is inside the atomic memory. After the coupling (\ref{coupl2}) and successful post-selection of the measured $X'_L$ quadrature in the interval $\langle-B,B\rangle$ around $X'_L=0$, the atomic mode is projected to the state with the following Wigner function
\begin{eqnarray}\label{wiga}
W'_{A}(x,p)=\frac{a\exp(-2x^2-2\frac{a^2p^2}{d^2})}{\pi(1+\exp(2x_0^2))d S}\left(\exp\left(\frac{2x_0^2a^2}{d^2}\right)\times\right.\nonumber\\
\left.\mbox{Cosh}\left[\frac{4\kappa ax_0x}{d}\right]\mbox{ERF}(p)+\mbox{CERF}(p)\right),\nonumber
\end{eqnarray}
where
\begin{eqnarray}
\mbox{ERF}(p)=\mbox{Erf}\left[\frac{\sqrt{2}}{a}\left(B-a\kappa p\right)\right]+ \mbox{Erf}\left[\frac{\sqrt{2}}{a}\left(B+a\kappa p\right)\right],\nonumber\\
\end{eqnarray}
and
\begin{eqnarray}
\mbox{CERF}(p)&=&\mbox{Re}\left[\mbox{Erf}\left[\frac{\sqrt{2}}{a}\left(ix_0a+B-\kappa p\right)\right]\right]+\nonumber\\
& &\mbox{Re}\left[\mbox{Erf}\left[\frac{\sqrt{2}}{a}\left(ix_0a+B+\kappa p\right)\right]\right].
\end{eqnarray}
The success rate of such the uploading process is
\begin{eqnarray}
S=\frac{1}{2(1+\exp(2x_0^2))}\left(2\exp(2x_0^2)\mbox{Erf}\left[\frac{\sqrt{2}B}{d}\right]+\right.\nonumber\\
\left.\mbox{Erf}[\frac{\sqrt{2}(ix_0a+B)}{d}]-\mbox{Erf}
[\frac{\sqrt{2}(ix_0a-B)}{d}]\right).\nonumber\\
\end{eqnarray}
If the squeezing post-correction $x\rightarrow xa/d$ and $p\rightarrow pd/a$ is applied, the Wigner function (\ref{wiga}) approaches the Eq.~(\ref{cat}) as $B$ goes close to zero, but with a reduced amplitude $x_0\rightarrow \kappa/d x_0$. If $B$ is enough small and $\kappa$ is sufficiently smaller than $a$ ($a/d\approx 1$) then the squeezing post-correction can be approximately omitted. But then also amplitude $x'_0=\kappa/d x_0$ of the states in the uploaded superposition decreases. Fortunately, this amplitude can be successfully compensated by a sufficient pre-squeezing ($a$ is small). Thus sufficiently large superposition of coherent states can be also uploaded into the quantum memory without any loss.

\begin{figure}
\centerline{\psfig{width=6cm,angle=0,file=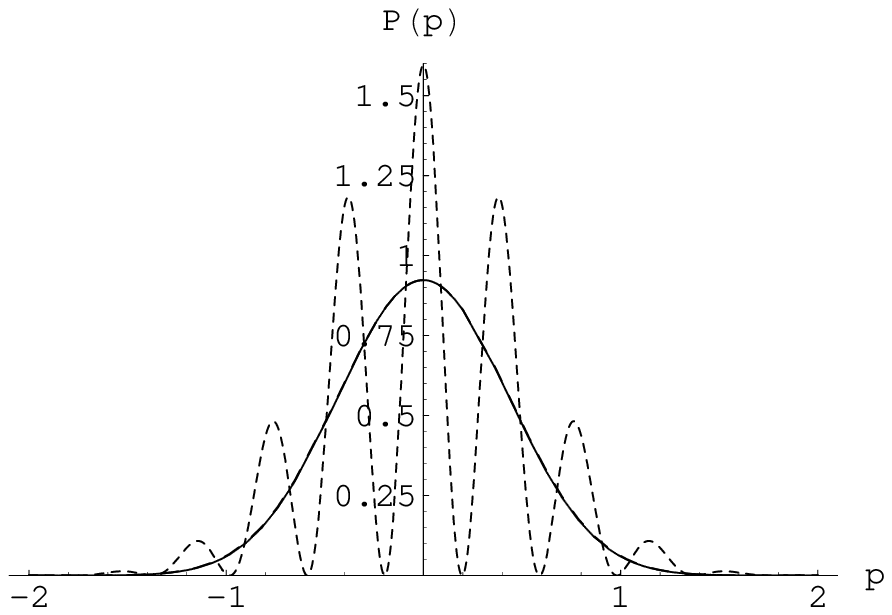}}
\centerline{\psfig{width=6cm,angle=0,file=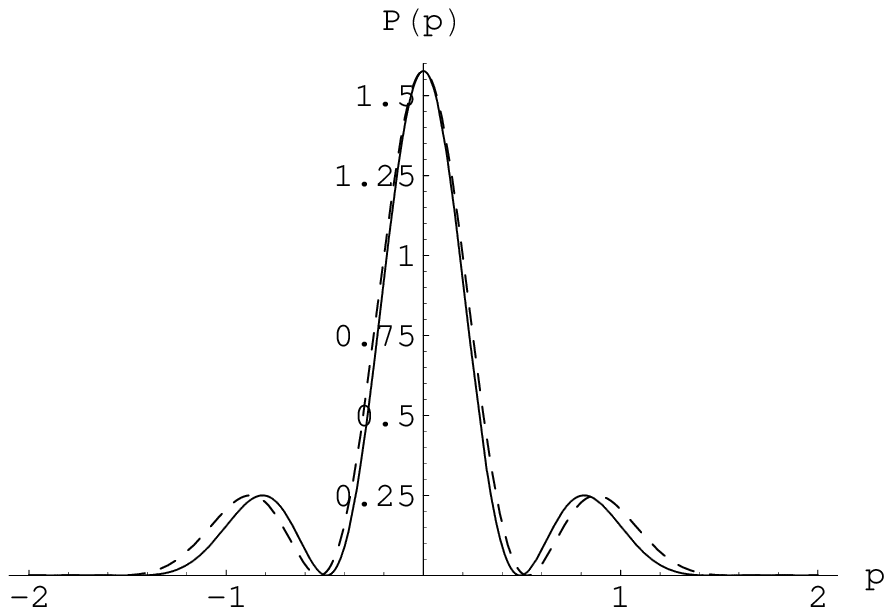}}
\caption{Interference in the marginal probability distribution of the superposition of coherent states uploaded to the memory with the pre-squeezing $a=1,0.25$ for $x_0=4$, $\kappa=0.1$, $B=0.01$. The uploaded superpositions have the amplitude $x_0=0.4,1.49$ with success rates $S=0.027,0.06$. The dotted line in the upper picture shows the initial interference of the coherent state superposition for $x_0=4$, in the lower picture it is ideal interference for $x_0=1.49$, for a comparison.}
\end{figure}

In a contrast to a simple phase-insensitive character of the single photon state, the superposition of coherent states exhibits more complex phase-space behavior. In one quadrature there two symmetrically located Gaussian peaks like for the classical mixture of the coherent states. On the other hand, the complementary quadrature shows multiple interference fringes covered by a Gaussian envelope. Because the fidelity averages both the interference as well as distinguishable peaks, it is not good measure of a quality of the uploaded superposition. From this reason, the marginal probability distribution
\begin{equation}
P(p)=\int_{-\infty}^{\infty}W'_A(x,p)dx
\end{equation}
exhibiting the interference is rather evaluated separately, to directly observe the interference in the superposition. The explicit form of marginal distribution
\begin{equation}
P(p)=\frac{a\exp(-2p^2)\left(\exp(2x_0^2)\mbox{ERF}(p)+\mbox{CERF}(p)
\right)}{Sd\sqrt{2\pi}(1+\exp(x_0^2))},\nonumber\\
\end{equation}
can be approximated by
\begin{equation}
P(p)\approx \frac{4\exp(-2\frac{d^2}{a^2}p^2)}{\sqrt{2\pi}\left(1+\exp\left(-2\frac{\kappa^2}{d^2}x_0^2\right)\right)}
\mbox{Cos}\left[\frac{2x_0\kappa}{a}p \right]
\end{equation}
for a small post-selection threshold $B$. Assuming $x_0=d/\kappa x'_0$, the final distribution is
\begin{equation}
P(p)\approx \frac{4\exp(-2\frac{d^2}{a^2}p^2)}{\sqrt{2\pi}\left(1+\exp\left(-2x^{'2}_0\right)\right)}
\mbox{Cos}\left[\frac{2dx'_0}{a}p \right].
\end{equation}
For $\kappa\ll a$ ($d/a\rightarrow 1$), $P(p)$ approaches the exact marginal distribution of the superposition of the coherent state only the amplitude $x'_0$ is reduced. As a result, the attenuation in the deterministic scheme, which for the coherent-state superposition introduces an excess noise in the uploaded state, is substituted here by only a reduction of $x_0$ without any additional noise. The uploaded state closely approaches the ideal pure state superposition with the reduced amplitude. The pre-squeezing plays here also a remarkable role. In Fig.~2 (upper picture), there is  a marginal distribution of the uploaded state without pre-squeezing. Evidently, the interference is vanishing. But if the sufficient pre-squeezing (-6dB) is applied (lower picture) then the fringes can be observed and also the success rate is higher. 

\section{Conclusion}

In Conclusion, the deterministic noise excess free record of unknown state and the probabilistic lossless upload of the resource state (single-photon state, superposition of coherent states) into quantum memory are proposed, both based on the original quantum memory experiment in Ref.~\cite{memory}. First, the deterministic scheme is able to record unknown quantum state only with a reduced transmission (maximally $1/4$). The pre-squeezing of the state of light going to the memory can compensate small coupling strength between light and atomic memory. But this record is without any trigger controlling the uploading process what is crucial, for example, for the upload of single photons.  Therefore, the probabilistic scheme which is able to upload single-photon state is proposed. It approaches unit fidelity of the upload at the cost of success rate, even for a weak coupling between the light and atoms. In this case, the pre-squeezing enhances the success rate of the upload. The same method can be used to upload the pure coherent-state superposition, at the cost of the success rate and reduced amplitude of the superposition, but with preservation of the purity of the superposition. In this case, the pre-squeezing not only enhances the success rate, but also effectively enlarges the reduced amplitude of the coherent states in the superposition and thus increases non-classicality of the uploaded state. Since the considered scheme has been used in the real experiment set-up, both the methods can be directly used to upload the non-classical resources into quantum memory. The proposed schemes are direct steps towards to storage and operations on highly non-classical quantum states in the continuous-variable quantum memory.


\medskip
\noindent {\bf Acknowledgments} This research has been supported by
projects  MSM 6198959213 and LC06007 of the Czech Ministry
of Education and grant 202/08/0224 of GA CR.
I also acknowledges a support by the Alexander von Humboldt
Foundation and EU grant FP7 212008 - COMPAS.




\begin{thebibliography}{99}

\bibitem{QM1}
C.H. van der Wal, M.D. Eisaman, A. Andr\' e, R. L. W.D.F. Phillips, A.S. Zibrov, and M.D. Lukin, Science 301, 196 (2003); D. N. Matsukevich and A. Kuzmich, Science 306, 663 (2004);

\bibitem{memory}
B. Julsgaard, J. Sherson, J.I. Cirac, J. Fiurasek, E.S. Polzik, Nature 432, 482 (2004).

\bibitem{Lee}
L.-M. Duan, G. Giedke, J. I. Cirac, and P. Zoller, Phys. Rev. Lett. 84, 2722 (2000);

\bibitem{Mandel}
P. Mandel, Opt. Lett. 4, 205 (1979).

\bibitem{Walls}
D.F. Walls and G.J. Milburn, Phys. Rev. A 31, 2403 (1985);

\bibitem{GG}
F. Grosshans, Phys. Rev. Lett. 94, 020504 (2005); M. Navascu\' es and A. Ac\' in, Phys. Rev. Lett. 94, 020505 (2005).

\bibitem{Opatrny05}
T. Opatrn\' y and J. Fiur\' a\v sek, Phys. Rev. Lett. 95, 053602 (2005).

\bibitem{Sherson06}
J. Sherson, A.S. Sorensen, J. Fiur\' a\v sek, K. Molmer and E.S. Polzik, Phys. Rev. A 74, 011802R (2006).

\bibitem{Muschik06}
Ch.A. Muschik, K. Hammerer, E.S. Polzik and J.I. Cirac, Phys. Rev. A 73, 062329 (2006).

\bibitem{Bruss}
in Lectures on Quantum Information (Section "Quantum Interface between Light and Atomic Ensembles"), eds. D. Bruss and G. Leuchs, Wiley-VCH, Berlin 2006.

\bibitem{logneg}
G. Vidal and R. F. Werner, Phys. Rev. A 65, 032314 (2002).

\bibitem{photon}
J. S. Neergaard-Nielsen, B. Melholt Nielsen, H. Takahashi, A. I. Vistnes, E. S. Polzik, Opt. Express 15, 7940 (2007).

\bibitem{Fiurasek}
J. Fiur\' a\v sek, R.-G. Patr\' on and N.J. Cerf, Phys. Rev. A 72, 033822 (2005).

\bibitem{Ou}
Z.Y. Ou, Quantum Semiclassic. Opt. 9, 599 (1997).

\bibitem{Lvovsky}
A.I. Lvovsky and J.H. Shapiro, Phys. Rev. A 65, 033830 (2002).

\bibitem{Ralph}
T. C. Ralph, A. Gilchrist, G. J. Milburn, W. J. Munro, and S. Glancy, Phys. Rev. A 68, 042319 (2003).

\end{thebibliography}
\end{document}